\documentclass[conference]{IEEEtran}
\usepackage[utf8]{inputenc}
\usepackage{newunicodechar}
\newunicodechar{⁠}{}
\usepackage{latexsym}
\usepackage{amssymb}
\usepackage{amsmath}
\usepackage[cmintegrals]{newtxmath}
\usepackage{bm}
\usepackage{multicol}
\usepackage{xcolor}
\usepackage{gensymb}
\usepackage{stfloats}
\usepackage{cite}
\usepackage[export]{adjustbox}
\usepackage{caption}
\usepackage{subcaption}
\usepackage{epstopdf}
\usepackage{amsfonts}
\usepackage{amsmath}
\usepackage[dvips]{epsfig}
\usepackage{lipsum,amsmath,multicol} 
\usepackage{svg}
\usepackage{booktabs}
\usepackage{siunitx}
\usepackage{hyperref}
\allowdisplaybreaks[4]	
\def\BibTeX{{\rm B\kern-.05em{\sc i\kern-.025em b}\kern-.08em
    T\kern-.1667em\lower.7ex\hbox{E}\kern-.125emX}}

\begin{document}
\title{Detection of sUAS in Urban Environments using Multi-Antenna Micro-Doppler Radar}

\author{
\IEEEauthorblockN{Chamindu Liyanage, ⁠Chirantha Kurukulasuriya, ⁠Chathuni  Wijegunawardana, ⁠Wikum Kumara, \\ Chamira U. S. Edussooriya}%
\IEEEauthorblockA{Department of Electronic and Telecommunication Engineering, University of Moratuwa, Moratuwa, Sri Lanka \\ chamira@uom.lk}\\
\vspace{-2mm}
\IEEEauthorblockN{Arjuna Madanayake}
\IEEEauthorblockA{Department of Electrical and Computer Engineering, Florida International University, Miami, FL, USA \\
Arcane AI and Wireless LLC, Miami, FL, USA\\
amadanay@fiu.edu}
}

\maketitle
\begin{abstract}
Sensing and early detection of small unmanned aerial systems (sUAS) are critically important in modern-day defense. In dense urban and indoor environments, detection becomes extremely challenging due to dense multipath, fading, low-altitude flight, and non-line-of-sight (NLOS) radio-frequency propagation. This paper presents a continuous-wave multiple-input multiple-output radar and a deep learning model for sUAS detection using NLOS signals. The radar operates at 2.47 GHz, and spectral correlation densities derived from rotational micro-Doppler signatures from the rotor blades are used as inputs to the deep learning model. Experimental results demonstrate an overall detection accuracy of $86.11\%$ across a dataset of five drone types, confirming the feasibility of sUAS detection in dense urban environments without direct line-of-sight conditions.
\end{abstract}
\begin{IEEEkeywords}
Small unmanned aerial vehicles, drones, micro-Doppler signatures, deep learning.
\vspace{-1ex}
\end{IEEEkeywords}

\section{Introduction}
\label{sec:int}

The radio spectrum is critical to defense. AI-enabled precision strike indoor–outdoor (PSIO) small unmanned aerial systems (sUAS), such as the ``Scorpio'', are reshaping urban warfare \cite{scorpio}. Defense against such AI-enabled PSIO technologies depends strongly on the ability to detect early and neutralize sUAS that are approaching at low altitudes and within highly dense built environments, including inside buildings. Yet RF countermeasures are designed for long-range, line-of-sight (LOS) detection and fail in the dense multipath and non-line-of-sight (NLOS) conditions of urban environments. In this paper, we explore multi-antenna RF sensing and detection of sUAS in dense multipath environments that are representative of urban built settings.

Recent approaches have addressed several aspects of radar based sUAS detection and localization. For instance, distributed multiple-input multiple-output (MIMO) continuous wave (CW) radar systems combined with cyclic spectral density have been explored for Doppler only localization \cite{yazici2024detection}. Other studies have incorporated contrastive learning with the Zhao–Atlas–Marks transform to improve robustness against multipath effects and environmental interference \cite{wu2022deep}. Addressing the detection of lowRCS plastic drones, Kim \textit{et al.} \cite{kim2024spectral} demonstrated that Spectral Kurtosis features offer superior discrimination between mechanical rotors and biological targets compared to standard spectral methods. Focusing on computational efficiency for edge deployment, Zhang \textit{et al.} \cite{zhang2025lightweight} developed RangeDopplerNet, a lightweight convolutional neural network (CNN) architecture that utilizes whitening layers to decorrelate microDoppler features while minimizing processing latency. Furthermore, Park \textit{et al.} \cite{park2025high} extended tracking capabilities into complex NLOS scenarios by employing a distributed MIMO frequency modulated continuous wave network that fuses local estimates via a weighted 2D MUSIC algorithm to resolve blind spots.

Prior work has also applied deep learning to RF radar detection using spectral correlation densities (SCD), including deep belief networks \cite{mendis2016deep}, long short-term memory networks \cite{sun2020micro}, CNN based approaches \cite{kim2016drone, scheiner2020seeing, he2022non}, and more recently wavelet transform
and YOLO~\cite{Gay2025}, and transformer-based architectures \cite{li2023nlost, caromi2021}. However, most of these methods primarily focus on LOS scenarios, and the problem of reliable sUAS detection in fully NLOS environments remains largely unaddressed.

This work targets practical NLOS sUAS detection using a hardware-validated CW MIMO Doppler radar system operating at 2.47 GHz with one transmitter and multiple receivers. To identify the presence of sUAS, we employ a deep learning-based detection framework that leverages the cyclostationary properties of radar returns. Specifically, SCD representations derived from target micro-Doppler signatures are used as input images to the model. 

By exploiting both SCD features and spatial diversity across receiver channels, the proposed approach improves robustness to multipath and clutter while outperforming single-channel sensing in complex urban and indoor environments. Experimental results on a dataset containing five sUAS types achieve an overall accuracy of $86.11\%$, demonstrating effective NLOS detection in dense urban environments. Furthermore, the system processes radar data in $2 s$ observation windows, while SCD generation and deep learning inference require only a few milliseconds, demonstrating its suitability for near real-time deployment. The paper demonstrates feasibility of MIMO sensing of UAS in dense multipath conditions in a typical indoor or built urban environment.

\section{MIMO CW Doppler Radar}
 
\subsection{Design Specifications}

\subsubsection{System Specifications}

The radar operates in the 2.4 GHz ISM band with 1 transmit channel and 4 receive channels. It is designed to detect sUAS at ranges up to 30 m, with a maximum transmitter power of 23 dBm  and a target signal-to-noise ratio (SNR) of 30 dB to ensure reliable micro-Doppler signature extraction.

The radar employs microstrip patch antenna array at 2.45~GHz to support compact, low-cost deployment. To achieve higher gain and a narrower beam, the patch antenna array is designed as a five-element series-fed linear array and Dolph--Chebyshev tapering \cite{dolph1946} applied to suppress sidelobe levels. The array fabricated using FR4 material resonates near 2.47~GHz with an impedance bandwidth of approximately 60~MHz, maintaining $|S_{11}| < -10$~dB over the 2.44--2.50~GHz band, a peak gain of approximately 4.1~dB, and a half-power beamwidth of about 21\textdegree{}. The directional radiation characteristics improve target illumination and reduce multipath interference, yielding more discriminative micro-Doppler features for deep learning–based NLOS detection.

\subsection{The MIMO CW Doppler Radar Transceiver}

\begin{figure*}[t!]
    \centering
    \includegraphics[width=0.97\textwidth]{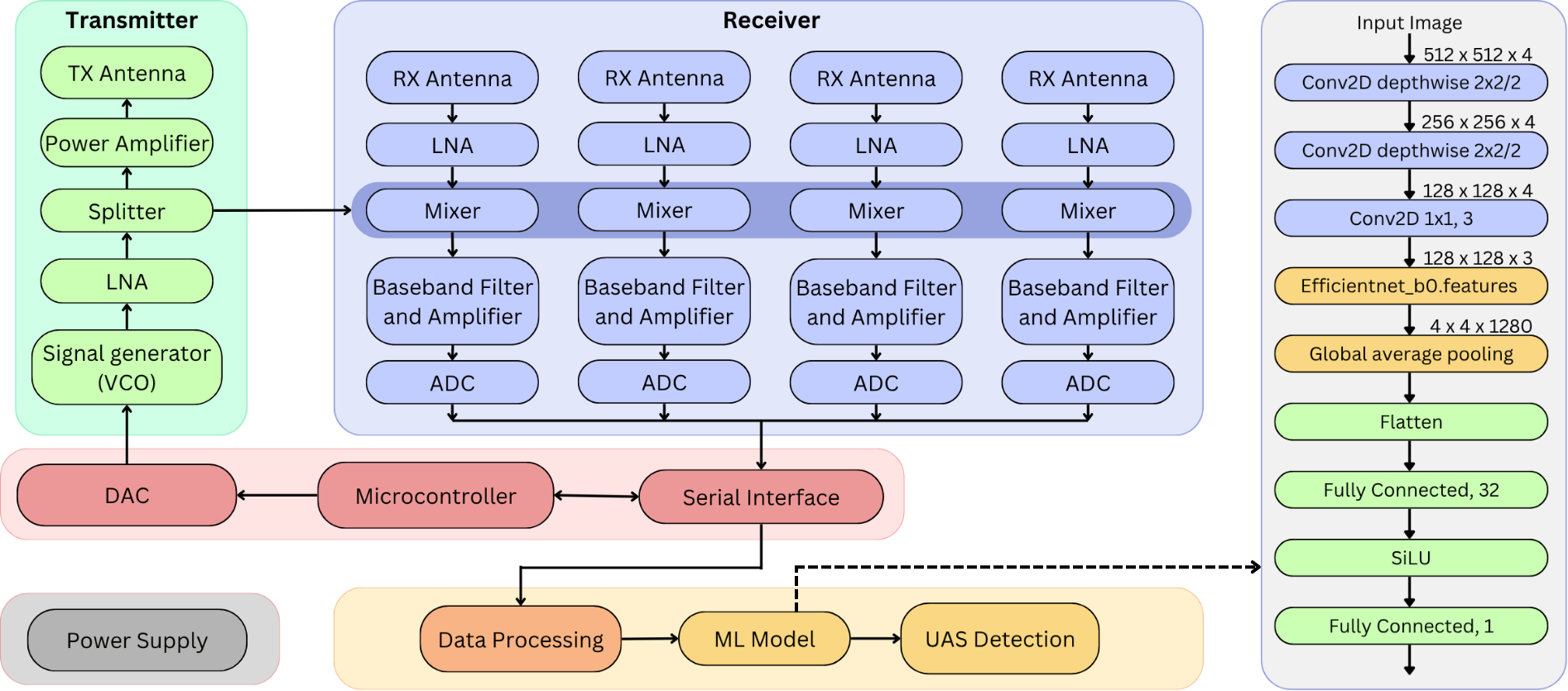}
    \caption{Transceiver Block Diagram of the proposed multi channel RF sensing system and the deep learning model. The transmitter generates a controlled RF signal using a VCO which is transmitted via the TX antenna, while a reference signal is provided to the receiver mixers for down-conversion. The receiver employs four receiver antennas. The digitized data are then processed and analyzed by an Efficientnet\_b0 based deep learning model for sUAS detection.}
    \label{fig:transceiver_block}
\end{figure*}

\begin{figure}[t]
    \centering
    \includegraphics[width=0.45\textwidth]{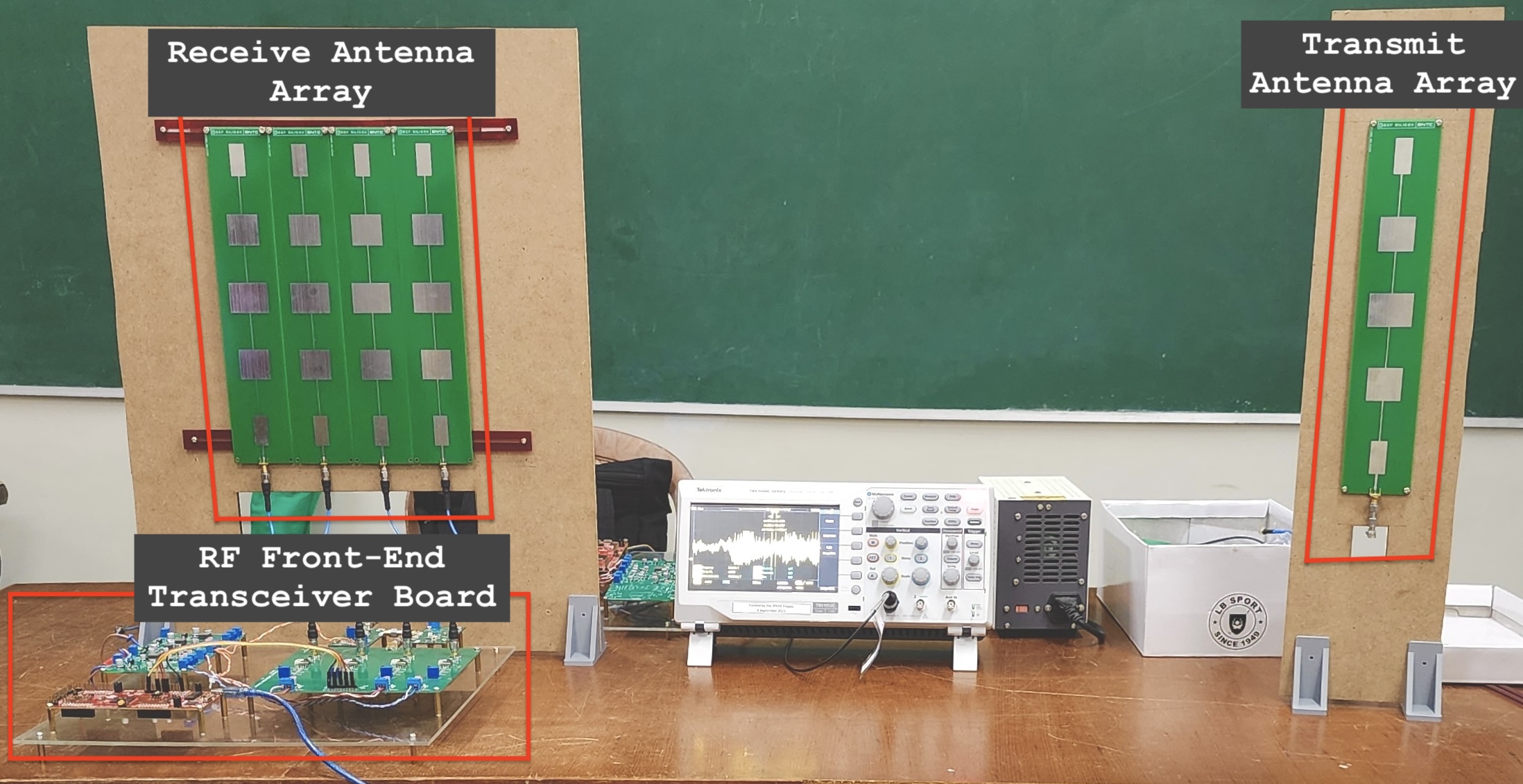}
    \caption{Photograph of the developed Doppler radar system showing the transceiver board and the microstrip antenna arrays used for transmission and reception.}
    \vspace{-10pt}
    \label{fig:transceiver_full}
\end{figure}

\begin{figure}[t]
    \centering
    \includegraphics[width=0.43\textwidth]{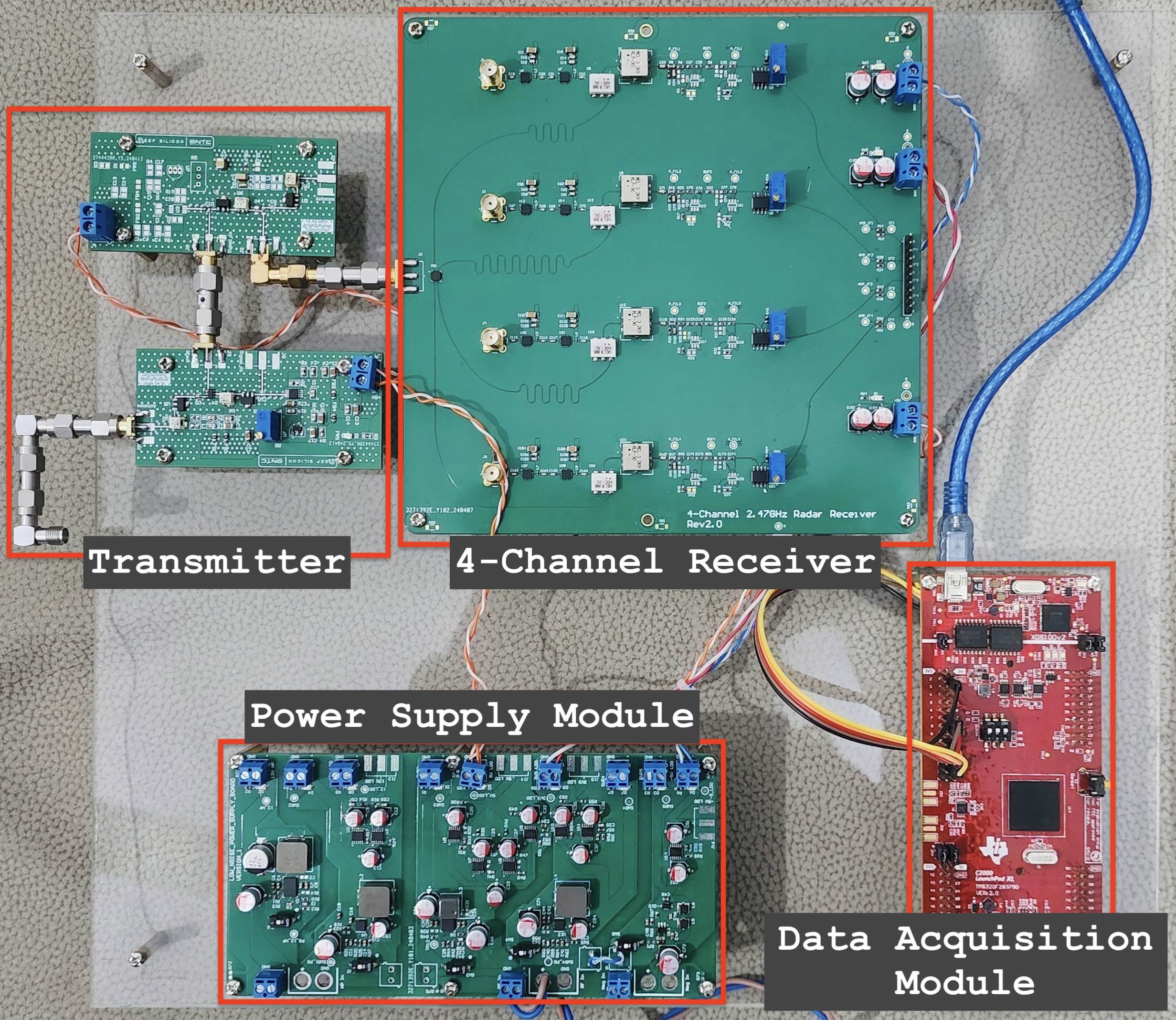}
    \caption{Implemented Doppler radar transceiver: transmitter (top left), four-channel receiver (top right), power supply (bottom left), and data acquisition module (bottom right).}
    \vspace{-10pt}
    \label{fig:pcb_implementation}
\end{figure}

The architecture of the Doppler radar transceiver is shown in Fig.~\ref{fig:transceiver_block}, implementation in Fig.~\ref{fig:transceiver_full}. Following ~\cite{mendis2016deep}, the system comprises three subsystems: transmitter, receiver, and data acquisition. The transmitter generates a CW signal at 2.45 GHz using a voltage-controlled oscillator (VCO), amplified by a low-noise amplifier (LNA) and power amplifier before transmission via a patch antenna. The receiver comprises four parallel channels, each consisting of a receive antenna, LNA, mixer, baseband filter, and analog-to-digital converter (ADC), which together extract micro-Doppler information from reflected signals. A Texas Instruments C2000 microcontroller digitizes the analog outputs and transmits the data to a host computer via USB for offline signal processing and classification. The microcontroller also provides a reference control voltage to the VCO for dynamic frequency tuning. The set of PCBs implemented are showed in Fig.~\ref{fig:pcb_implementation}.

\vspace{-5pt}
\section{Proposed NLOS Detection Approach}
\label{sec:nlos}

\subsection{Generation of Spectral Correlation Densities}
\label{sec:scf}
Micro-Doppler radar returns from rotating UAS propellers exhibit cyclostationary behavior due to periodic modulation induced by blade rotation. These periodicities give rise to spectral correlations at specific cyclic frequencies that can be captured using SCD analysis \cite{gardner2006cyclostationarity}. Prior work has demonstrated the effectiveness of SCDs for UAS detection under LOS conditions \cite{mendis2016deep}. Here, we hypothesize that SCDs remain effective in cluttered and NLOS environments, as the cyclostationary features associated with propeller motion are intrinsically linked to source periodicity rather than to the propagation channel. As established by Gardner \textit{et al.} \cite{gardner2006cyclostationarity}, periodic signal-generating mechanisms induce cyclostationary second-order statistics that manifest as spectral correlations at specific cyclic frequencies. Since stationary clutter and multipath propagation predominantly introduce non-cyclostationary or weakly cyclostationary components, these cyclic features are expected to be less sensitive to environmental effects than conventional power spectral features.

Cyclostationary micro-Doppler features are extracted using the FFT Accumulation Method (FAM) \cite{roberts1991fam}. For each measurement, a $10s$ raw complex baseband signal is collected from all four receiver channels, yielding an input tensor of size $(4, 10000)$ at a sampling rate of 1 kHz. To increase the number of training samples and capture temporal variation, the raw signal is segmented into six overlapping patches of 2048 samples. Each segment is normalized and passed through a Butterworth lowpass filter with a cutoff frequency of 180 Hz to suppress high-frequency noise. The SCDs are then computed independently for each receiver channel using FAM, producing an SCD representation of size $(4,512,512)$ per input sample.

\subsection{Deep Learning Model}
\label{sec:dl}

The SCD tensor of shape $(4,512,512)$ serves as input to the deep learning model, which is trained to detect sUAS in both LOS and NLOS environments. Vision-based architectures were considered, and CNNs and transformer-based architectures \cite{vaswani2017attention} were compared. Transformer-based models such as vision transformers~\cite{dosovitskiy2020image} capture long-range dependencies effectively but require substantially larger datasets and higher computational resources. CNN-based models, by contrast, offer an inherent inductive bias and efficiently capture local spatial features. Given the dataset's size constraints, a CNN-based approach was selected for its balance of accuracy, efficiency, and deployment feasibility.

\subsubsection{Model Architecture} 
Several CNN architectures were evaluated as the deep learning backbone, including EfficientNet \cite{tan2019efficientnet}, VGG-19 \cite{simonyan2014VGG}, and custom CNN designs. EfficientNet-B0 was selected as the feature extractor, owing to its strong performance and lightweight design. Two depthwise convolutions with a stride of 2 downsample the input tensor, followed by a pointwise convolution to reduce channel depth, before passing to the feature extractor. The full architecture is shown in the left side of the Fig.~\ref{fig:transceiver_block}.

\subsubsection{Implementation Details}
Given the limited data availability, augmentation techniques including random sampling and additive Gaussian noise with power varying from 0.1 to 0.6 were employed. Custom layer weights were randomly initialized, whilst the EfficientNet backbone was initialized with ImageNet~\cite{deng2009imagenet} weights.

To mitigate class imbalance, binary cross-entropy loss was weighted by a factor $\alpha$ for the positive class, calculated from the training set class distribution, yielding $\alpha = 0.282$:
\begin{equation} \label{eq:bceloss_adjusted}
    L_{BCE} = -\frac{1}{N}\sum_{i=1}^{N}\alpha \cdot y_i\cdot \log p(y_i) + (1-y_i)\cdot \log (1 - p(y_i))
\end{equation}

Training proceeded in two stages. In the first stage, only the custom layer weights were trained with EfficientNet frozen for 70 epochs, using a learning rate of $5\times10^{-4}$ and weight decay of $1\times10^{-5}$. In the second stage, all parameters including the EfficientNet backbone were unfrozen for a further 70 epochs, with the learning rate reduced to $1\times10^{-5}$ and weight decay increased to $5\times10^{-4}$ to reduce overfitting.

\section{Experimental Results}
\label{sec:experiments}

\begin{figure}[t]
    \centering
    \includegraphics[width=0.45\textwidth]{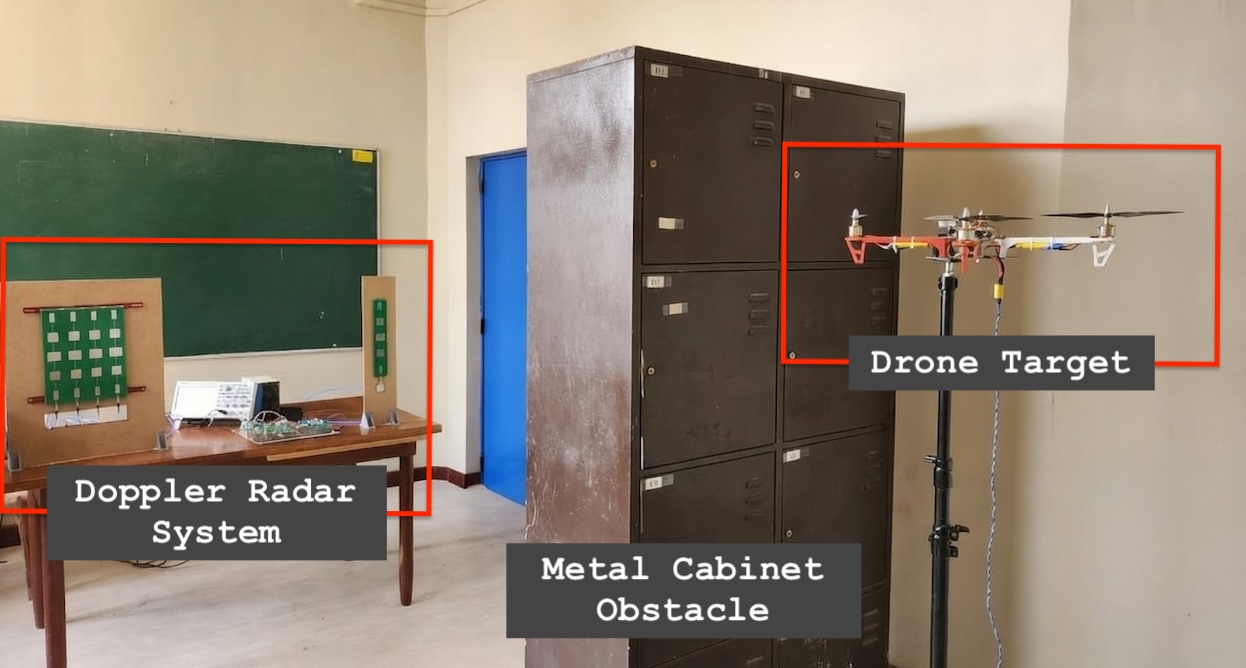}
    \caption{Indoor NLOS measurement setup illustrating the Doppler radar system and a drone target, where a metallic cabinet is introduced to block the direct line-of-sight path.}
    \vspace{-10pt}
    \label{fig:experimental_setup}
\end{figure}

\begin{figure*}[t!]
    \centering
    \includegraphics[width=0.85\textwidth]{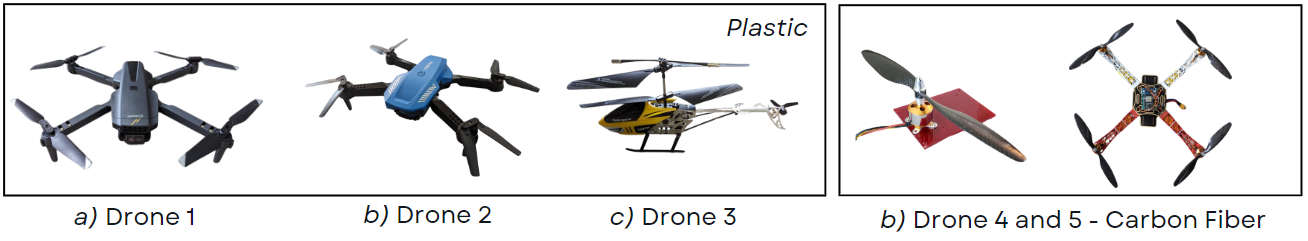}
    \caption{Small unmanned aerial systems (sUAS) used to construct the dataset, including platforms with plastic propellers (Drones 1–3) and carbon-fiber propellers (Drones 4–5).}
    \label{fig:drone types}
\end{figure*}

\begin{figure*}[t!]
    \centering
    \includegraphics[width=0.9\textwidth]{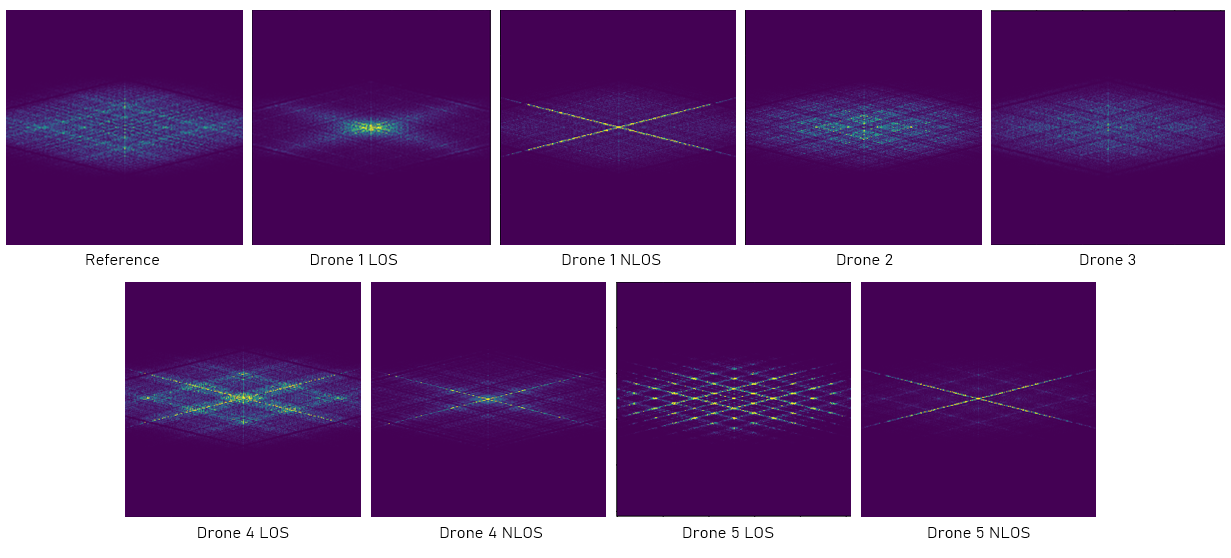}
    \caption{SCD ${S_x(f,\alpha)}$ representations of different sUAS under LOS and NLOS conditions, with spectral frequency $f$ on the horizontal axis and cyclic frequency $\alpha$ on the vertical axis.}
    \label{fig:scd_patterns}
\end{figure*}

\subsection{Experimental Setup}
Data acquisition was performed under both LOS and NLOS conditions using the constructed hardware. Considering both conditions is essential, as the system is designed to operate across a wide range of real-world scenarios including direct LOS propagation, partially obstructed links, and fully NLOS signal paths. Training on both conditions enables the model to learn the effects of obstruction and multipath, yielding a more robust system that generalizes well to diverse operating environments. Experiments were conducted at multiple indoor and outdoor locations to maximize dataset variability. Indoor experiments took place in laboratory and living room settings; outdoor experiments were carried out in open and semi-urban surroundings. For NLOS measurements, metal cabinets, doors, and walls were used to obstruct the direct path between the drone and the transceiver. For both conditions, the UAS was kept stationary, and only the propeller rotation was varied to validate the proposed hypothesis. A representative setup is shown in Fig.~\ref{fig:experimental_setup}.

Five sUAS types, shown in Fig.~\ref{fig:drone types}, were used to construct the dataset. The first three feature plastic propellers; the remaining two use carbon-fiber propellers. Each sUAS produces a unique SCD pattern characterized by specific, repeating peaks at particular frequency–cyclic frequency pairs. Reference (non-sUAS) data exhibits no SCD peaks or structured pattern, highlighting the discriminative nature of the sUAS-induced cyclostationary features.

Comparing LOS and NLOS SCD patterns for the same sUAS, identical cyclostationary peaks and structured spectral patterns are observed in both cases, confirming that propeller-induced cyclostationary characteristics are preserved under NLOS conditions. Under NLOS, the overall SCD magnitude is reduced due to multipath propagation, particularly affecting higher Doppler and higher-order cyclic components. Representative SCD patterns are shown in Fig.~\ref{fig:scd_patterns}.

\subsection{Detection Results}
Detection performance for each sUAS type under LOS and NLOS conditions is presented in \autoref{table:detection_rates}. Drones 1–3, whose propellers are constructed from plastic, exhibit comparatively lower detection rates. Drone 3 achieves the lowest detection rate, attributable to its smaller physical size and reduced radar cross-section. Drones 4 and 5 achieve higher detection rates in both environments; drone 5 exceeds a detection rate of 0.95 in both scenarios, demonstrating strong robustness to occlusion.

The final model achieves an overall classification accuracy of $86.11\%$ on the test dataset. In practical sUAS detection systems, recall, precision, and F1-score are critical metrics, as missed detections pose significant security risks. The model attains a recall of $87\%$ for the sUAS class, indicating a high probability of identifying true targets, alongside a precision of $81\%$ and an F1-score of $84\%$, demonstrating a favorable balance between detection sensitivity and false alarm control. The model correctly classifies $80\%$ of non-sUAS instances.

\begin{table}[t!]
\caption{Accuracy of detecting different types of sUAS under LOS and NLOS conditions.}
\label{table:detection_rates}
\centering
\begin{tabular}{ p{2cm}|p{2cm}|p{2cm}  } 
    \hline
    \textbf{Drone type} & \textbf{LOS detection rate} & \textbf{NLOS detection rate} \\ [1ex] 
    \hline
    Drone 1 & $0.85$ & $0.81$ \\ 
    Drone 2 & $0.79$ & $-$ \\ 
    Drone 3 & $0.60$ & $-$ \\ 
    Drone 4 & $0.86$ & $0.82$ \\ 
    Drone 5 & $0.97$ & $0.95$ \\ 
    \hline 
\end{tabular}
\end{table}

\subsection{Comparison with Single Channel Inputs}
To assess the benefit of multiple receivers, an equivalent model was constructed with the first few layers modified to accept a single-channel input. Reducing the SCD input from four channels to one requires replacing the first three layers with the convolution layers shown in \autoref{table:conv_layers}; all other layers, including the EfficientNet backbone, are kept identical to enable a fair comparison. The single-channel model achieves an accuracy of $73.77\%$, more than $10\%$ below that of the four-channel model, confirming the importance of spatial diversity for NLOS detection.

\begin{table}[t!]
\caption{Convolutional Layers for Single Channel Input Model}
\label{table:conv_layers}
\centering
\begin{tabular}{ p{3cm}|p{1cm}|p{1cm}|p{1.2cm}  } 
    \hline
    \textbf{Layer Name} & \textbf{Kernel Size} & \textbf{Stride} & \textbf{Output Channels}   \\ [1ex] 
    \hline
    Depthwise\_Convolution\_1 & $2 \times 2$ & $2$ & $1$ \\ 
    Pointwise\_Convolution\_1 & $1 \times 1$ & $1$ & $2$ \\ 
    Depthwise\_Convolution\_2 & $2 \times 2$ & $2$ & $2$ \\ 
    Pointwise\_Convolution\_2 & $1 \times 1$ & $1$ & $3$ \\ 
    \hline 
\end{tabular}
\end{table}

\section{Conclusion and Future Work}
This paper presents a multi-antenna Doppler radar system for detecting sUAS in cluttered indoor and NLOS environments. By leveraging cyclostationary micro-Doppler features through SCD analysis and a lightweight CNN based deep learning architecture, the proposed system demonstrates robust detection under severe multipath conditions. The four channel MIMO configuration significantly outperforms a single-channel baseline, achieving an overall accuracy of $86.11\%$ and confirming the importance of spatial diversity in NLOS sensing. The model performs strongly for sUAS equipped with carbon-fiber propellers in both LOS and NLOS conditions, benefiting from their stronger and more consistent micro-Doppler returns. Reduced performance is observed for UAVs with plastic propellers, primarily owing to their lower radar cross-section and the limited transmit power imposed by regulatory constraints.

\normalsize
\bibliographystyle{IEEEtran}    
\bibliography{IEEEabrv,refs}

\end{document}